\documentclass[preprintnumbers,showpacs,prl,twocolumn]{revtex4}
\usepackage{amsmath}
\usepackage[dvips]{texdraw}
\usepackage{amssymb}
\usepackage{mathrsfs}

\input{tcilatex}

\begin{document}

\title{Local Average and Scaling Property of $1/f^{\alpha }$ Random Fields}
\author{Hai Lin}
\affiliation{Jadwin Hall, Princeton University, Princeton, NJ 08544}

\begin{abstract}
We use the techniques developed in \cite{van} to study the local average of random
fields with spectral density $1/f^{\alpha }$. We study their scaling
properties and show that the self-similarity of $1/f$ random fields is
preserved under the local average. We study how the original random fields
can be recovered from the locally averaged ones. We also study the
derivative of the locally averaged random fields as a way to get the
spectral density. Finally, we propose the generalization of local average by
means of an arbitrary response function.
\end{abstract}

\maketitle



\section{Introduction}

Random fields are associated with probabilities. If we want to study a
one-dimensional random field \cite{van}\ $X(t),$\ we would like to specify
not only the probability density function $P(x,t)$\ for this field but also
the joint probability density functions $P(x_{1},t_{1};x_{2},t_{2})$,..., $%
P(x_{1},t_{1};x_{2},t_{2},...,x_{n},t_{n})$ and so on. Here the capital
letter $X$\ denotes the random field, and the lower case letter $x$\ denotes
the possible value it can take, and $t$\ is the time. With those
probabilities in hand, we can study many statistical properties of the
random fields. Some are first order effects, like the mean or expectation
value $E[X(t)]$=$\int\limits_{-\infty }^{+\infty }xP(x,t)dx$, some are
second order effects, like the 2-point correlation function and covariance,
and some are higher orders, like the n-point correlation functions. The n-th
order effects are related to the n-point joint probability density functions.

Most random fields have "memories", namely they have correlations of their
values taken at different times. Some don't have correlations and are thus
indeed random. Those with non-zero correlations are not that random. Their
covariance can be studied as deterministic properties. The covariance $%
B(t_{1},t_{2})$\ is defined as the correlation of the fluctuation around the
mean, and is associated with the joint probability functions:%
\begin{eqnarray}
&&B(t_{1},t_{2})  \notag \\
&=&E[(X(t_{1})-E[X(t_{1})])(X(t_{2})-E[X(t_{2})])] \\
&=&\int\limits_{-\infty }^{+\infty
}x_{1}x_{2}[P(x_{1},t_{1};x_{2},t_{2})-P(x_{1},t_{1})P(x_{2},t_{2})]dx_{1}dx_{2}
\notag
\end{eqnarray}

We can see that the covariance is non-zero only if the 2-point joint
probability density function is not simply the product of two probability
functions, namely there is some entanglement between the two events $%
(x_{1},t_{1})$\ and $(x_{2},t_{2})$\ such that the conditional probability
density function is not the same with the single probability density
function: $P(x_{1},t_{1}\mid x_{2},t_{2})\neq P(x_{1},t_{1}),$\ and $%
P(x_{2},t_{2}\mid x_{1},t_{1})\neq P(x_{2},t_{2}).$\ 

If the random field is stationary, $B(t_{1},t_{2})$\ is only a function of $%
\tau =t_{1}-t_{2}:$\ $B(t_{1},t_{2})=B(\tau ).$\ The covariance function $%
B(\tau )$\ can be more conveniently studied by using the Fourier transform.
The Fourier transform of the covariance function is the spectral density $%
S(\omega ).$ Without loss of generality, we can assume that the means of the
random fields are zero, since the means are deterministic and can be
subtracted for simplicity of discussion. The random field $X(t)$\ can be
decomposed into Fourier components $F(\omega )$: 
\begin{equation}
X(t)=\func{Re}\int\limits_{-\infty }^{+\infty }F(\omega )e^{i\omega t}d\omega
\end{equation}%
and thus the covariance of $X(t)$\ becomes the Fourier transform of the
square of \ $F(\omega ):$ 
\begin{equation}
E[X(t_{1})X(t_{2})]=\func{Re}\int\limits_{-\infty }^{+\infty }\langle
F^{2}(\omega )\rangle e^{i\omega (t_{1}-t_{2})}d\omega
\end{equation}%
Since $F(\omega )$\ is the amplitude of each Fourier mode, the square of it
should be considered as the power or energy spectral density:%
\begin{equation}
S(\omega )=\langle F^{2}(\omega )\rangle
\end{equation}%
So we have the Wiener-Khinchine relation between $B(\tau )$\ and$\ S(\omega
) $\cite{van}$:$ 
\begin{equation}
B(\tau )=\int\limits_{-\infty }^{+\infty }S(\omega )\cos \omega \tau d\omega
\end{equation}

A class of random fields has the spectral density of the form ($\alpha $ is
some positive constant): 
\begin{equation}
S(\omega )\sim \frac{1}{\omega ^{\alpha }}.\ 
\end{equation}%
They are very common in nature, especially for the case $\alpha =1,$\ the
1/f noise, which is reported to be ubiquitous \cite{review}. The 1/f noise
almost present everywhere in every kinds of phenomena, \ from resistors \cite%
{voss} to traffic \cite{traffic}, from brain to number theory \cite{number},%
\cite{number2}, etc. The microscopic origin of 1/f noise is widely studied,
in e.g. \cite{review},\cite{Kau}-\cite{Zanella},\cite{tang1},\cite{tang}.

In real experiment when we measure the value of a random field, we often
measure an average value in a small domain. The idea of the local average 
\cite{van} of the random fields is thus very practical and important. In
this paper we will study the local average of the random field with spectral
density 1/$\omega ^{\alpha },$ and their scaling properties under local
average. We will also discuss the derivative\ and the inverse process/
inverse problem of the local average.

I will first discuss their scaling properties in the covariance functions,
both before and after local average. Since the spectral density is
proportional to $\frac{1}{\omega ^{\alpha }},$ we have 
\begin{eqnarray}
B(\tau ) &\sim &\int\limits_{0}^{+\infty }\frac{1}{\omega ^{\alpha }}\cos
\omega \tau d\omega \  \\
&=&\tau ^{\alpha -1}\int\limits_{0}^{+\infty }\frac{1}{(\omega \tau
)^{\alpha }}\cos \omega \tau d(\omega \tau ).
\end{eqnarray}
So we see 
\begin{eqnarray}
B(\tau ) &\sim &\tau ^{\alpha -1},\ \qquad (\mathrm{for}\qquad \alpha >0),
\label{alpha big 0} \\
B(\tau ) &\sim &\delta (\tau ),\qquad (\mathrm{for}\qquad \alpha =0)
\label{alpha zero}
\end{eqnarray}

We can see that the exponent $\alpha $\ determines how much correlated the $%
\frac{1}{\omega ^{\alpha }}$\ random fields are. The bigger the $\alpha ,$\
the more correlated it is.

If $\alpha >1,$\ $B(\tau )$\ increases with $\tau ,$\ it is positively
correlated. For example, in the Brownian motion, we know from the Einstein
relation that the covariance of the position of the Brownian particle
increases linearly with time ($D$\ is the coefficient of diffusion): $%
E[X(t_{1})X(t_{2})]=2D(t_{1}-t_{2}),$\ clearly $B(\tau )\sim \tau ,$\ so for
Brownian motion, $\alpha =2$.

If $\alpha <1,$\ $B(\tau )$\ decreases with $\tau ,$\ it is negatively
correlated. If $\alpha =1,$\ it's the 1/f noise, in which $B(\tau )\sim 
\mathrm{const}.$\ There is no a frequency scale or time scale in ideal 1/f
noise, since the covariance function is constant everywhere, and after a
rescaling of the time:$\tau \rightarrow n\tau ,$\ it remains the same: $%
B(n\tau )=B(\tau ).$\ 

If $\alpha =0,$\ it's the white noise and its spectral density $S(\omega )$\ 
$\sim \frac{1}{\omega ^{0}}$\ is constant and $B(\tau )\sim \delta (\tau )$.

Generally, the random fields with $\frac{1}{\omega ^{\alpha }}$\ spectral
density are self-similar. The curve $x$\ vs. $t$\ of a realization of these
kind of random fields looks like a fractal, e.g. \cite{book}-\cite{music2}:
For general $\alpha ,$\ under a rescaling:$\tau \rightarrow n\tau ,$\ $%
B(n\tau )=n^{\alpha -1}B(\tau ),$namely when we magnify $\tau $\ $n$\ times
and simultaneously multiply the random field by a factor $n^{(1-\alpha )/2}$%
, the curve $x$\ vs. $\tau $\ looks the same. In other words, they are
fractals with fractal dimensions $d=2+(1-\alpha )/2$ \cite{dimension},\cite%
{fractal dimension}.

However, these scaling dependence should be appropriate in the large $\tau $%
\ regime. In small $\tau $, perturbation shows $B(\tau )$\ is a Gaussian
with maximum $B(0)=\sigma _{X}^{2}(t).$\ 

Now let's come to the local average of the random fields with spectral
density $\frac{1}{\omega ^{\alpha }}$.

\section{Local Average of $1/f^{\protect\alpha }$ Random Fields}

The local average\ \cite{van} of $\ $the random field $X(t)$\ with a window
of length $T$\ is defined as 
\begin{equation}
X_{T}(t)=\ \frac{1}{T}\int\limits_{t-\frac{T}{2}}^{t+\frac{T}{2}%
}X(t_{1})dt_{1}.  \label{local average def}
\end{equation}

There could be many ways to calculate the relations of the covariance
functions and spectral densities between the averaged and unaveraged random
fields.

One way is by defining a variance function $\gamma (T)=\frac{\sigma
_{X_{T}}^{2}}{\sigma _{X}^{2}}.$\ It is shown \cite{van} that $\gamma (T)$\
is a response to the covariance function:%
\begin{equation}
\gamma (T)\sim \frac{2}{T}\int\limits_{0}^{T}(1-\frac{\tau }{T})B(\tau )d\tau
\end{equation}%
For $S(\omega )\sim \frac{1}{\omega ^{\alpha }},$\ if we use\ (\ref{alpha
big 0}) $B(\tau )\sim \tau ^{\alpha -1},$\ we have $\gamma (T)\sim T^{\alpha
-1}.$\ 

Another way is by analyzing the spectral density. It is shown in \cite{van}
that: 
\begin{equation}
S_{X_{T}}(\omega )=[\frac{\sin (\omega T/2)}{\omega T/2}]^{2}S_{X}(\omega ).
\label{spectral average}
\end{equation}%
So we have 
\begin{equation}
B_{X_{T}}(\tau )\sim \int\limits_{0}^{+\infty }[\frac{\sin (\omega T/2)}{%
\omega T/2}]^{2}\frac{1}{\omega ^{\alpha }}\cos \omega \tau d\omega .
\end{equation}

When $T$\ is very small it's clear that $[\frac{\sin (\omega T/2)}{\omega T/2%
}]^{2}\approx 1,$\ so both the spectral density and covariance function
before and after the local average are almost the same. $S_{X_{T}}(\omega
)\approx S_{X}(\omega ),B_{X_{T}}(\tau )\approx B(\tau ).$\ It remains a $%
\frac{1}{\omega ^{\alpha }}$\ spectral density when $T$\ is small.

When $T$\ is not small, consider the function: 
\begin{equation}
\sin ^{2}(\omega T/2)\cos \omega \tau =\frac{1}{2}[\cos \omega \tau -\frac{1%
}{2}\cos \omega (\tau +T)-\frac{1}{2}\cos \omega (\tau -T)].\ 
\end{equation}%
So we have 
\begin{eqnarray}
&&B_{X_{T}}(\tau )  \notag \\
&\sim &\frac{1}{T^{2}}\int_{0}^{+\infty }[2\cos \omega \tau -\cos \omega
(\tau +T)  \notag \\
&&-\cos \omega (\tau -T)]\frac{1}{\omega ^{\alpha +2}}d\omega , \\
\ &\sim &\frac{1}{T^{2}}[2\tau ^{\alpha +1}-(\tau +T)^{\alpha
+1}-(\left\vert \tau -T\right\vert )^{\alpha +1}]\ 
\end{eqnarray}

When $\alpha =1,B_{X_{T}}(\tau )$ $\sim \mathrm{const}.$\ The local average
of $1/\omega $\ noise is still a $1/\omega $\ noise. This is the
manifestation of the self-similarity of $1/\omega $\ noise.

For $\alpha \neq 1,$when $T$\ is small, we can expand $T$\ around $\tau $\
in the above function, $B_{X_{T}}(\tau )$$\sim \tau ^{\alpha -1},$\ which is
the same as the unaveraged one.

If $T$ is big, we can expand $\tau $\ around $T,$\ then%
\begin{equation}
B_{X_{T}}(\tau )\sim \frac{1}{T^{2}}[2\tau ^{\alpha +1}-2T^{\alpha
+1}-\alpha (\alpha +1)\tau ^{2}T^{\alpha -1}].  \label{covariance before der}
\end{equation}

\section{Derivatives of Locally Averaged Random Fields}

The spectral density of the local averaged random field can be achieved in
another way in which we first calculate the spectral density of the
derivative of the local averaged field and then use the relation $S_{\overset%
{.}{X}_{T}}(\omega )=\omega ^{2}S_{X_{T}}(\omega )$.

The derivative of the local averaged random field $X_{T}=\frac{1}{T}%
\int\limits_{t-\frac{T}{2}}^{t+\frac{T}{2}}X(t_{1})dt_{1}$\ is:%
\begin{equation}
\overset{.}{X}_{T}=\frac{dX_{T}}{dt}=\ \frac{1}{T}[X(t+\frac{T}{2})-X(t-%
\frac{T}{2})]  \label{x der}
\end{equation}

The covariance of $\overset{.}{X}_{T}$ can be known from the information of
the covariance of $X,$\ since%
\begin{eqnarray}
&&E[\overset{.}{X}_{T}(t)\overset{.}{X}_{T}(t+\tau )]  \notag \\
&=&\frac{1}{T^{2}}\{E[X(t+\frac{T}{2})X(t+\frac{T}{2}+\tau )]  \notag \\
&&+E[X(t-\frac{T}{2})X(t-\frac{T}{2}+\tau )]  \notag \\
&&\ -E[X(t-\frac{T}{2})X(t+\frac{T}{2}+\tau )]  \notag \\
&&-E[X(t+\frac{T}{2})X(t-\frac{T}{2}+\tau )]\}
\end{eqnarray}

If $X(t)$\ is stationary, the above equation amounts to%
\begin{equation}
B_{_{\overset{.}{X}_{T}}}(\tau )=\frac{1}{T^{2}}[2B_{_{X}}(\tau
)-B_{_{X}}(\tau +T)-B_{_{X}}(\tau -T)]
\end{equation}

If we use $B(\tau )\sim \tau ^{\alpha -1},$ for $\alpha >0,$ then\ 
\begin{equation}
B_{_{\overset{.}{X}_{T}}}(\tau )\sim \frac{1}{T^{2}}[2\tau ^{\alpha
-1}-(\tau +T)^{\alpha -1}-(\left\vert \tau -T\right\vert )^{\alpha -1}]
\label{covariance after der}
\end{equation}

It's interesting to compare the covariance function of the local average of
the $\frac{1}{\omega ^{\alpha }}$ random fields before and after taking a
derivative, i.e. eqn (\ref{covariance before der}) and (\ref{covariance
after der}).

Making a Fourier transform of both sides of\ (\ref{covariance after der}),%
\begin{eqnarray}
S_{\overset{.}{X_{T}}}(\omega ) &=&\frac{1}{T^{2}}[2S_{X}(\omega
)-e^{-i\omega T}S_{X}(\omega )-e^{i\omega T}S_{X}(\omega )]\   \notag \\
&=&\frac{4}{T^{2}}S_{X}(\omega )\sin ^{2}(\omega T/2)
\end{eqnarray}

On another hand , we have $S_{\overset{.}{X}_{T}}(\omega )=\omega
^{2}S_{X_{T}}(\omega ),$\ so we get%
\begin{equation}
S_{X_{T}}(\omega )=[\frac{\sin (\omega T/2)}{\omega T/2}]^{2}S_{X}(\omega )
\end{equation}%
which is in agreement with eqn (\ref{spectral average}).

\section{Inverse Process of Local Average}

One may ask the question whether we can get the information of the original
random field $X(t)$\ from the information of the local averaged one $X_{T}(t)
$. If we have two locally averaged fields with very close lengths of the
window, ie, we measure $X_{T_{1}}(t)$\ and $X_{T_{2}}(t)$\ in which $T_{1}$\
and $T_{2}$\ are very close to each other (suppose $T_{1}>T_{2}$\ ), since $%
X_{T_{1}}(t)$\ and $X_{T_{2}}(t)$\ are integral over range $T_{1}$\ and $%
T_{2},$\ their difference reveals the information of $X_{(T_{1}-T_{2})}(t).$%
\ 
\begin{eqnarray}
&&T_{1}X_{T_{1}}(t)-T_{2}X_{T_{2}}(t)  \notag \\
&=&\frac{T_{1}-T_{2}}{2}[X_{(T_{1}-T_{2})}(t-\frac{T_{1}+T_{2}}{2})  \notag
\\
&&+X_{(T_{1}-T_{2})}(t+\frac{T_{1}+T_{2}}{2})]
\end{eqnarray}

By making the auto-covariance of both sides, we can get the cross-covariance
of $X_{T_{1}}(t)$\ and $X_{T_{2}}(t)$.

If $T_{1}$\ and $T_{2}$\ are so close that we can make a derivative with
respect to $T,$\ the above equation becomes:%
\begin{equation}
X_{T}(t)+T\frac{\partial X_{T}(t)}{\partial T}=\frac{1}{2}[X(t-\frac{T}{2}%
)+X(t+\frac{T}{2})].
\end{equation}%
Together with eqn (\ref{x der})$,$\ we may solve $X(t)$\ in terms of $%
X_{T}(t):$\ 
\begin{equation}
X(t)=X_{T}(t+\frac{T}{2})+T\frac{\partial X_{T}(t+\frac{T}{2})}{\partial T}-%
\frac{T}{2}\frac{\partial X_{T}(t+\frac{T}{2})}{\partial t}  \label{ahead}
\end{equation}%
or 
\begin{equation}
X(t)=X_{T}(t-\frac{T}{2})+T\frac{\partial X_{T}(t-\frac{T}{2})}{\partial T}+%
\frac{T}{2}\frac{\partial X_{T}(t-\frac{T}{2})}{\partial t}  \label{delay}
\end{equation}

This recovers the original random field $X(t)$ from the locally-averaged one 
$X_{T}(t)$. The two expressions (\ref{ahead}),(\ref{delay}) are consistent.
The derivative with respect to $T$\ is understood as to compare two local
averages whose lengths of windows are very close to each other.

The derivative of a local average is the local average of the derivative: 
\begin{eqnarray}
\frac{dX_{T}(t)}{dt} &=&\ \frac{1}{T}[X(t+\frac{T}{2})-X(t-\frac{T}{2})] \\
&=&\frac{1}{T}\int_{t-\frac{T}{2}}^{t+\frac{T}{2}}\overset{.}{X}%
(t_{1})dt_{1}=\left( \overset{.}{X}(t)\right) _{T}
\end{eqnarray}

\section{Generalized Local Average and Response Function}

Local average defined in \cite{van} can be generalized to weighted local
average, characterized by a normalized function $h(t)$:%
\begin{equation}
X_{h}(t)=\ \int_{-\infty }^{+\infty }X(t_{1})h(t_{1}-t)dt_{1},
\end{equation}%
where \ $\int_{-\infty }^{+\infty }h(t_{1}-t)dt_{1}=1.$

In this language, the ordinary local average (\ref{local average def}) in 
\cite{van} is the case where the function $h(t_{1}-t)$\ is a step function: 
\begin{eqnarray}
h(t_{1}-t)\ &=&\frac{1}{T},\ \qquad \mathrm{for}\qquad \left\vert
t_{1}-t\right\vert \leqslant \frac{T}{2}, \\
h(t_{1}-t)\ &=&0,\ \qquad \mathrm{for}\qquad \left\vert t_{1}-t\right\vert >%
\frac{T}{2}.
\end{eqnarray}

The generalization of $h(t)$ to other functions might be useful when the
actual random field we measure is an uneven average, namely it is possible
that in the averaging process when we measure in the real experiments, the
values near the center has bigger weights than those away from the center,
or inversely. The general $h$ respects this experimental detail. For
example, perhaps $h(t_{1}-t)$\ could be a Gaussian centered at $t.$

In this case, the spectral density of the weighted local average is 
\begin{equation}
S_{X_{h}}(\omega )=\left\vert H(\omega )\right\vert ^{2}S_{X}(\omega ),
\end{equation}%
where $H(\omega )$\ is the Fourier transform of the function $h(t_{1}-t).$

We see that this actually means that the local average is a response to the
original unaveraged random field by the response function $h$.


\begin{acknowledgments}
I am very grateful to Prof. VanMarcke for helpful discussions.
\end{acknowledgments}

\end{document}